
\input phyzzx
\overfullrule=0pt

\def\p{\partial}

\rightline {UTTG-31-92}
\rightline {December 1992}
\bigskip\bigskip
\title{Classical Solutions in Two-Dimensional String Theory
and Gravitational Collapse
\foot{Work supported in part by NSF grant PHY 9009850 and
R.~A.~Welch Foundation.}}

\vskip 3cm
\author{Jorge G. Russo,\foot{e-mail: russo@utaphy.ph.utexas.edu } }
\address {Theory Group, Department of Physics, University of
Texas\break
Austin, TX 78712}
\vskip 2cm

\abstract
\singlespace
A general solution to the $D=2$ 1-loop beta functions equations
including tachyonic back reaction on the metric is presented.
Dynamical black hole (classical) solutions
representing  gravitational collapse of tachyons are
constructed. A discussion on the correspondence
with the matrix-model approach is given.
\vfill\endpage

\Ref\bh {J.A. Harvey and A. Strominger, Chicago preprint, EFI-92-41;
J.G. Russo, L. Susskind and L. Thorlacius, Phys.Rev. D46 (1992) 3444.}

\Ref\mm {D. Gross and N. Milkovic, Phys. Lett. 238B (1990) 217; E.
Brezin, V. Kazakov and Al. B. Zamolodchikov, Nucl. Phys. B338 (1990) 673;
G. Parisi, Phys. Lett. 238B (1990) 209; P. Ginsparg and J. Zinn-Justin,
Phys. Lett. 240B (1990) 333.}

\Ref\dj {S.R. Das and A. Jevicki, Mod. Phys. Lett. A5 (1990) 1639.}

\Ref\POLCO{J. Polchinski, Nucl. Phys. B346 (1990) 253.}

 \Ref\gk{D. Gross and I. Klebanov, Nucl. Phys. B352 (1990) 671
and Nucl. Phys. B359 (1991) 3;
 A.M. Sengupta and S.R. Wadia, Int. J. Mod. Phys. A6
 (1991) 1961.}

\Ref\djr{K. Demeterfi, A. Jevicki and J.P. Rodrigues, Nucl. Phys. B362
(1991) 173 and Nucl. Phys. B365 (1991) 199;
G. Mandal, A.M. Sengupta and S.R. Wadia, Mod. Phys. Lett. A6
(1991) 1465;
G. Moore, Nucl. Phys. B368 (1992) 557; G. Moore, M.R. Plesser
and S. Ramgoolam, Nucl. Phys. B377 (1992) 143.}

 \Ref\pol {J. Polchinski, Nucl. Phys. B362 (1991) 125.}

\REF\witten{E.~Witten, IAS preprint, IASSNS-HEP-92/24.}

\REF\dasdahr{S. R. Das, Tata Institute preprint, TIFR-TH-92/62 (1992);
A. Dhar, G. Mandal and S.R. Wadia, Tata Institute preprint, TIFR-TH-92/63
(1992). }

\Ref\djbh{J.G. Russo, Texas preprint, UTTG-27-92.}

\Ref\muell{M. Mueller, Nucl. Phys. B337 (1990) 37;
S. Elitzur, A. Forge and E. Rabinovici, Nucl.Phys.B359 (1991) 581;
G. Mandal, A. Sengupta and S.R. Wadia, Mod. Phys. Lett. A6
(1991) 1685.}

\Ref\wittbh{E. Witten, Phys. Rev. D44 (1991) 314.}

\Ref\DVV{R. Dijkgraaf, E. Verlinde and H. Verlinde, Nucl. Phys. B371
(1992) 269. }

\Ref\ark{A.A. Tseytlin, Phys. Lett. B268 (1991) 175.}

\Ref\lykk {S.P. De Alwis and J. Lykken, Phys. Lett.B269 (1991) 264.}

\Ref\dfk {P. Di Francesco and D. Kutasov, Phys.Lett. B261 (1991) 385.}

Recently there has been an increasing interest in understanding
the quantum mechanical paradoxes associated with Hawking radiation
in the context of simple two-dimensional models containing
black holes (see ref. [\bh ] and references therein).
The knowledge of exact results coming from discrete approaches
[\mm-\pol ] also converts two-dimensional string theory in a
simple laboratory to
explore the quantum physics underlying black hole geometries
(for recent discussions in this direction see refs.[\witten-\djbh ]).
The black hole interpretation of some classical solutions of
$D=2$ string theory [\muell ] was discovered in ref. [\wittbh ]
by using an exact conformal field theory constructed from
a gauged WZW model.

In physical situations the black hole should appear by gravitational
collapse of the only propagating mode of the $D=2$ critical string
theory, which at low energies is $\eta\equiv e^{-\phi}T $, where
$\phi $ is the dilaton and $T$ is the tachyon.
In ref. [\djbh ] it was observed that there is a natural
effective geometry which emerges in
scattering processes in the Das-Jevicki collective field theory [\dj ]
which for a distant observer (far away from the horizon)
 appears as a black hole configuration.
But in view of the simplicity of the exact classical S-matrix
derived in the matrix model approach
one may be concerned about the actual existence of dynamically formed
black holes.
The fact that the static black hole survives
after $\alpha' $ corrections,
as found in refs. [\DVV ,\ark ], is encouraging,
but unfortunately it is not enough to prove that
there is gravitational collapse in the model.
Note that `shock-wave' configurations
constructed by a simple matching of the `exact' static metric of
ref. [\DVV ] with the linear dilaton vacuum
cannot be implemented without an exact account for
all higher powers of $T$.
Here we will explicitly construct solutions representing
transitions between the linear dilaton vacuum and the static
black hole solutions by
dynamical gravitational collapse of tachyons.
The essential assumption involved in these solutions is that
part of the incoming energy is not reflected. As discussed
below, if a pulse is instead fully reflected there is a transitory
period during which there is a black-hole type configuration, but the
final state is the linear dilaton vacuum. The resulting picture is
in agreement with the matrix-model approach.

To leading $\alpha '$ order, the tree-level string effective action for
the metric, dilaton and tachyon is given by,
$$
S = \int d^2x \sqrt{-G}
\bigl[e^{-2\phi}(R+4\p_\mu\phi\p^\mu \phi+c
-\p_\mu T\p^\mu T+{2\over \alpha '}T^2+O(T^3)] \ \ ,
\eqn\eff
$$
where $c=-8/\alpha' $.

The equations of motion, neglecting $O(T^3)$ terms in
the action \eff , are given by
(henceforth we set $\alpha'=2$)
$$
R_{\mu\nu}+2D_\mu\p _\nu \phi=\p_\mu T\p_\nu T \ ,
\eqn\uno $$
$$ 4 (\p\phi)^2-2\nabla^2\phi -T^2-4=0\ , \eqn\dos$$
$$
\nabla^2 T -2\p_\mu\phi\p^\mu T + T=0\ \ .
\eqn\tres
$$
If $T=0$ the general solution is the Witten black hole.
Around this background the field $\eta=e^{-\phi }T $ becomes
massless far away from the black hole.
The static solution to this equations with nonvanishing
static $T$ was found in ref. [\lykk ].

It is convenient to work in the gauge in which the dilaton
is linear, $\phi=-x $, and the metric has the diagonal form
$$ds^2=Adt^2+Bdx^2 \ \ .\eqn\metr $$
Then the 01, 00, 11 components of eq. \uno \  and the
dilaton equation \dos\  take the form
$${\dot B\over B}=\dot TT'  ,
\eqn\unoa $$
$${A\over 2}R-{A'\over B}=\dot T^2 \ ,
\eqn\unob
$$
$${B\over 2}R+{B'\over B}=T'T'\ , \eqn\unoc
$$

$$4(1-B)-(\log|B/A|)'=BT^2 \ \eqn\unod $$
(as conventional primes and dots denote derivatives with respect
to $x$ and $t$ respectively). In this gauge the curvature is
$$R=-{1\over \Delta}[{d\over dx}({A'\over\Delta}) +
{d\over dt }({\dot B\over \Delta})]
\ ,\ \ \ \Delta=\sqrt{-AB} \ \ .  \eqn\RRR $$
{}From eq. \unoa \ we get
$$B=f(x)\exp[\int^t dt \dot TT'  ]\ \ .
\eqn\BBB$$
Combining eqs. \unob\ and \unoc \ we find
$$(\log |AB|)' =T'T'-{B\over A}\dot T^2\ \ \ ,
\eqn\logAB
$$
from where it follows
$$
AB=-g(t)\exp[\int^x dx (T'T'-{B\over A}\dot T^2)]\ \ .
\eqn\AB $$
The arbitrary function $g(t)$ can be set to 1 by a time
reparametrisation.

It is worth noting that if the tachyon self-interactions are
nonderivative, then eqs.\BBB \  and \AB \ are exact to all order
in powers of $T$. This is because eq.\uno\ is not modified by
nonderivative tachyon self-interactions (but of course it is
modified by two (sigma model) loop corrections).

Let us now assume that $A=-1+O(e^{-2x})\ ,\ \ B=1+O(e^{-2x})$ and
$T=e^{-x}\eta =O(e^{-x})$.
Then, to this
$T^2$ order, we can set $B/A\cong -1 $ in
eq. \AB \ obtaining
$$AB=-\exp[\int^x dx (T'T'+\dot T^2)]\ \ \ .
\eqn\ABB $$
In order to determine the arbitrary function $f(x)$
we insert eq.\BBB \ into the dilaton equation \unod . We
get $f'=2(1-f)$, i.e. $f(x)=1+Me^{-2x}\ ,\ \ M=$const.,
and the compatibility condition
$$T^2-\dot T^2 +2\int^tdt\  \dot T T'' +4\int^t dt\ T'\dot T=0\ \ .
\eqn\TTTT
$$
Since to the lowest order we have
$$ T''-\ddot T +2T'+T=0 \ , \eqn\linT
$$
the condition \TTTT \ is automatically satisfied on-shell (as guaranteed
by covariance). To see this explicitly multiply  eq.\linT \
by $\dot T$ and integrate over $t$.

If we set $T=0$ we recover the Witten black hole solution where
$M$ is the ADM mass of the black hole. Here we will be interested in
black holes purely formed by gravitational collapse so we shall set
$M$ to zero.

To the present $O(T^2)$ approximation the solution can thus be written as
$$ds^2=-(1+\int^x_\infty dx(T'T'+\dot T^2) -\int^t_{-\infty}
dt \ \dot TT')dt^2
+(1+ \int^t_{-\infty}dt \ \dot TT')dx^2\ \ .
\eqn\solmet
$$
The simplest example is that in which $T$ is a static
configuration $T=\mu e^{-x}$, $\mu $=const. (or $\eta=\mu $).
In this case the solution takes the form
$$ds^2=-(1-{1\over 2}\mu^2e^{-2x})dt^2+dx^2\ ,\ \ \phi=-x\ \ \ ,
\eqn\static $$
which is similar to the Witten black hole solution but not exactly
the same (in the linear dilaton gauge Witten metric is different).

Now consider the more physical case of a dynamical
gravitational collapse. Let $T$ be an incoming localized wave packet,
for example
$$\eta(x)=\eta(x^+)={a\over \cosh (2x^+)}\  ,\ \ \ x^+=x+t\ \ ,
\eqn\pak $$
or $T=a {e^{-x}\over \cosh(2x^+ )}$. Then
$$
\int _{-\infty }^t dt \dot TT'= {2\over 3}a^2 e^{-2x}
\big( 1+\tanh(2x^+)-{\sinh(2x^+)\over\cosh^3(2x^+)}-
{3\over 4}{1\over\cosh^2(2x^+)}\big)\ \ ,
\eqn\ttt $$
$$\int _{\infty}^x dx (T'T'+\dot T^2) = -2a^2e^{2t}\bigg(
-{3\over 4i}\log{1-iz\over 1+iz}
-{7\over 2}{z\over z^2+1}+{22\over 3}{z\over (z^2+1)^2}
-{16\over 3} {z\over (z^2+1)^3}\bigg)\ \ ,
\eqn\xxx $$
$$z\equiv e^{-2x^+}\ \ .$$
There is an expanding horizon
given by the equation $x=x_h(t)$,
$$1+\int^{x_h}_\infty dx(T'T'+\dot T^2)
-\int^t_{-\infty}dt \ \dot TT'=0\ \ .
\eqn\hor
$$
Let us now consider late times,  $t\to\infty $.
The leading terms in eq.\xxx\ , $O(e^{-2x^+})$ , cancel out, so
$\lim _{t\to\infty} \int _{\infty}^x dx (T'T'+\dot T^2) =O(e^{-4x-2t})
\cong 0$. Therefore the final metric has the form
$$ds^2=-(1-me^{-2x})dt^2 +(1-me^{-2x})^{-1}dx^2 \ \ ,
\ \ \ m={4\over 3} a^2\ .\eqn\final$$
This is nothing but the Witten static black hole solution.
Thus the solution given by eqs.\solmet ,\ttt , \xxx \
describes the transition from
the linear dilaton vacuum to the static black hole solution
due to collapsing matter.

The essential assumption involved in the above solution is that
$\eta $ has only ingoing component. More generally,
if the outgoing component carries less energy than the ingoing component
then the final metric will be given by eq. \final \ with $m$ representing
the energy that was not reflected.
Whether all the energy is reflected or not is something that
will be dictated by the higher
order terms in powers of $T$ which become important in the
region $x^-=t-x\to\infty $. Let us see that
if we assume that part of the energy is not reflected back then
the solution \solmet \
implies that the final state is a black hole.
This is not trivial, since
this solution is only valid in the region $x>>x_h$, i.e.
in the region $1>>me^{-2x}$. The question is, then, whether a  horizon
really forms in dynamical processes
or the black hole geometry is just an illusory
large-distance effect.
In the region $x^+\to\pm \infty $ the tachyon goes to zero
so we can exactly solve the differential equations by
using the results of refs. [\DVV, \ark]. Therefore the exact final metric
which al large distances matches with \final \ is just the DVV metric
(up to three-loop ambiguities pointed out in ref. [\ark ])
with ADM mass equal to $m$,
\foot {The coordinates $(t,q)$ are `canonical' coordinates in
the Das-Jevicki field theory (in these coordinates the
kinetic term of the bosonic hamiltonian has the canonical form). In this
theory an effective metric given by eq.(24)  appears in a natural way
(see ref. [\djbh ]). The coordinate $q$ is related to the radial
coordinate $r$
of ref. [\DVV] by $\cosh (q) =\sqrt{ 1+m}\cosh (r)$.}
$$ds^2=A_0(q)dt^2-A_0^{-1}(q)dq^2\ \ ,\ \ \ \
A_0(q)\equiv 1-m{1\over \sinh^2(q) }\ ,
 \eqn\exact$$
$$\phi =-{1\over 2}\log \sinh(2q) +\phi_0=-x\ \ ,\ \ \
\phi_0\equiv{1\over 4}\log {4(m+1)\over m} $$
$$R=-A_0''(q)={4m\over\sinh^4(q)}(\cosh^2(q)+{1\over 2}) $$
This metric  has an event horizon at $x={1\over 2}\log m$
and a singularity at $x=-\infty$.

However, the scattering amplitudes for low-energy waves in
the matrix-model approach show that everything
is reflected off the wall. This seems to be the case also
in the present continuum
formalism. Indeed, by plugging the solution \solmet \ into the
tachyon equation of motion we see that it has the structure
$$ T''-\ddot T +2T'+T=O_V(T^2) +O(T^3)\ , \eqn\TTT
$$
where $O_V(T^2)\sim e^{-2x}$ contains only a tachyon
self-interaction contribution, coming from the ``$T^3$" term in
the lagrangian. That is, the gravitational part
only influences the next order $O(e^{-3x})$.
But it is well known that the on-shell S-matrix of low-energy tachyons
due to the $O_V(T^2)$ term  is unitary (see e.g. refs. [\POLCO , \dfk ]).
Therefore, in accordance with the prediction from the matrix-model
approach,
an ingoing low-energy wave should undergo total reflection, since
it will first feel the $T^3$ interaction, being the
gravitational forces exponentially suppressed.

Let us consider the following example:
$$\eta (x)={a\over \cosh(2x^+) }+{a\over \cosh(2x^-) }
\eqn\refle
$$
Eq. \refle \  represents a pulse that is fully reflected.
It is straightforward to compute metric \solmet\ in this case
to see that the final state is the linear dilaton vacuum.
There is a transitory period during which the large-distance geometry
looks like a black hole configuration.
Consider now the case of an incoming `step' energy-density
pulse (a constant source of ingoing energy-density which turns on
at some time). Even when
everything is reflected, the final stationary geometry will be
of the form \final \  due to the constant presence of
energy in the space.
The picture is in fact very similar to the picture discussed
in ref. [\djbh] in the context of Das-Jevicki collective field theory.
In the view of a distant observer,
there is a horizon which appears to be
beyond the reflection point (this is the physical boundary of the space;
the wall is not rigid, but moves in and out as the system evolves).
Any nontrivial geometry produced by localized pulses disappears
after the pulses have been reflected and get away from the wall, etc.
\foot {A four-dimensional analog would be a spherical shell of
collapsing matter which, before reaching the Schwarzchild radius,
bounces and expands due to self-interactions.}

To conclude, the equivalence of the continuum and discrete approaches
implies that a low-energy pulse is completely reflected and
hence there is not actual formation of event horizons.
Eq.\solmet \ describes the evolution of the large-distance geometry
in terms of the source.
Transitory black-hole type configurations, with
virtual horizons and singularities appearing beyond the physical
boundary of the space, are present in generic physical processes.
The present description breaks down for high-energy pulses which
can cross the barrier.

\bigskip
 \noindent $\underline {\rm Acknowledgements}$: The author is grateful
 to W. Fischler and  A. Tseytlin for useful discussions.

\refout
\end